\documentclass[preprintnumbers, floatfix, letterpaper, twocolumn,aps,prd,epsfig,nofootinbib,natbib,longbibliography]{revtex4-1}
\usepackage{graphicx}
\usepackage{epstopdf}
\usepackage{latexsym}
\usepackage{amssymb}
\usepackage{amsmath}
\usepackage{color}
\usepackage{mathrsfs}
\usepackage{xparse}
\usepackage{bbding}
\usepackage{pifont}
\usepackage{comment}
\usepackage{ulem}
\usepackage{float}
\usepackage[inline]{enumitem}
\usepackage{soul}
\usepackage[caption=false]{subfig}

\let\Right\right
\let\Left\left
\makeatletter
\def\right#1{\Right#1\@ifnextchar){\!\right}{}}
\def\left#1{\Left#1\@ifnextchar({\!\left}{}}
\makeatother

\usepackage[
            pdfstartview=FitH,
            bookmarksnumbered=true,
            bookmarksopen=true,
            colorlinks,
            linkcolor=blue,
            anchorcolor=green,
            citecolor=blue
            ]{hyperref}
\begin{document}

  \renewcommand\arraystretch{2}
 \newcommand{\bq}{\begin{equation}}
 \newcommand{\eq}{\end{equation}}
 \newcommand{\bqn}{\begin{eqnarray}}
 \newcommand{\eqn}{\end{eqnarray}}
 \newcommand{\nb}{\nonumber}
 \newcommand{\lb}{\label}
 
\newcommand{\La}{\Lambda}
\newcommand{\va}{\scriptscriptstyle}
\newcommand{\be}{\nopagebreak[3]\begin{equation}}
\newcommand{\ee}{\end{equation}}

\newcommand{\ba}{\nopagebreak[3]\begin{eqnarray}}
\newcommand{\ea}{\end{eqnarray}}

\newcommand{\la}{\label}
\newcommand{\n}{\nonumber}

\newcommand{\R}{\mathbb{R}}

 \newcommand{\cb}{\color{blue}}
    \newcommand{\cc}{\color{cyan}}
        \newcommand{\cm}{\color{magenta}}
\newcommand{\rc}{\rho^{\scriptscriptstyle{\mathrm{I}}}_c}
\newcommand{\rd}{\rho^{\scriptscriptstyle{\mathrm{II}}}_c} 
\NewDocumentCommand{\evalat}{sO{\big}mm}{%
  \IfBooleanTF{#1}
   {\mleft. #3 \mright|_{#4}}
   {#3#2|_{#4}}%
}
\newcommand{\PRL}{Phys. Rev. Lett.}
\newcommand{\PL}{Phys. Lett.}
\newcommand{\PR}{Phys. Rev.}
\newcommand{\CQG}{Class. Quantum Grav.}

\title{Linear instability of hairy black holes in Horndeski theory}

\author{Chao Zhang${}^{a, b}$}
\email{ phyzc@cup.edu.cn}

\author{Tao Zhu${}^{c, d}$}
\email{zhut05@zjut.edu.cn; Corresponding author}

\affiliation{${}^{a}$Basic Research Center for Energy Interdisciplinary, College of Science, China University of Petroleum-Beijing, Beijing 102249, China\\
${}^{b}$Beijing Key Laboratory of Optical Detection Technology for Oil and Gas, China University of Petroleum-Beijing, Beijing, 102249, China\\
${}^{c}$Institute for theoretical physics and cosmology, Zhejiang University of Technology, Hangzhou, 310023, China\\
${}^{d}$United Center for Gravitational Wave Physics, Zhejiang University of Technology, Hangzhou, 310023, China}

\date{\today}

\begin{abstract}

The Horndeski theory gives the most general model of scalar-tensor theories. It draws a lot of attentions in recent years on its black holes, celestial dynamics, stability analysis, etc. It is important to notice that, for certain subclasses of Horndeski models, one can obtain analytic solutions to the background fields. This provides us with a good opportunity to investigate the corresponding stability problems in details. Specially, we may find out the constraints to the model or theory, under which the stability conditions can be satisfied. 
In this paper, we focus on a subclass of the Horndeski theory and a set of analytic background solutions are considered. On top of that, the odd-parity gravitational perturbation and the 2nd-order Lagrangian are investigated. 
With careful analysis, the instability is identified within the neighborhood of event horizon. We are thus able to exclude a specific geometry for the model. 
It is interesting to notice that, such an instability is implanted in the structure of the corresponding Lagrangian, and will not by erased by simply adding numerical constraints on the coupling parameters. 
As a starting point of our research, this current work provides insights into further exploration of Horndeski theories.

\end{abstract}


\maketitle
\section{Introduction}

\renewcommand{\theequation}{1.\arabic{equation}} \setcounter{equation}{0}

The detection of the first gravitational wave (GW) from the coalescence of two massive black holes (BHs) by advanced LIGO/Virgo marked the beginning of a new era --- {\it the GW astronomy} \cite{Ref1}. After 100 years, one of Einstein's crucial predictions was finally confirmed \cite{Einstein1916}. Following this observation, about 90 GW events have been identified by the LIGO/Virgo/KAGRA (LVK) scientific collaborations (see, e.g., \cite{GWs, GWs19a, GWs19b, GWsO3b}). In the future, more advanced ground- and space-based GW detectors will be constructed  \cite{Moore2015, Gong:2021gvw},  such as Cosmic Explorer \cite{CE} and the Einstein Telescope \cite{ET}, LISA \cite{LISA}, TianQin \cite{Liu2020, Shi2019}, Taiji \cite{Taiji2}, and DECIGO \cite{DECIGO}. These detectors will enable us to probe signals with a much wider frequency band and larger distances. This triggered the interest in the observation of, e.g.,  quasi-normal mode (QNM) of black holes \cite{Chao2023a},  extreme mass ratio inspirals (EMRIs) \cite{Gong2023}, etc. 

GWs emitted during the ringdown stage are in general studied with the purterbation theory \cite{Berti18, Chandra92}.
In general relativity (GR) it has been studied extensively, including scalar, vector, and tensor (gravitational) perturbations \cite{Iyer1987}. 
In fact, the QNMs that generated from gravitational perturbations are closely related to the test and confinement of theories of gravity \cite{Zack2020}. 
On the other hand, the resultant QNM frequencies reflect some aspects of the stability of the spacetime under consideration \cite{Berti2009}. At the same time,  the gravitational perturbations themselves onto the background fields also play an important role in the stability analysis.

For a spherically symmetric geometry, such a perturbation problem could be divided into odd- and even-parity sectors, with the latter for most cases has a more intricate structure (see., e.g., \cite{Fang2023}). 
It is important to notice that, under the framework of GR, the characters of gravitational perturbations are relatively easy to track (see., e.g., \cite{Chao2021}). However, moving to the realm of modified theories of gravity, the structure of the original action (or equivalently, the Lagrangian) can be quite sophisticated,  which often set barriers to us in achieving desired physical information from it. 
One way to tackle this problem is to work on top of the original perturbed Lagrangian and eliminate its non-dynamical terms (according to the degrees of freedom of the theory), before processing to further analysis \cite{Chao2023a}.  

Basically, we can substitute all the background as well as perturbation terms into the original Lagrangian. Treating a perturbation term as a 1st-order infinitesimal quantity, we can extract the 2nd-order part from such a Lagrangian. After that, by using effective mathematical techniques and (probably) introducing suitable gauge-invariants, one can (in principle) dramatically simplify that and eliminate all the non-dynamical terms. It is then straightforward to manage the reduced Lagrangian. 
Clearly, such a reduced Lagrangian is crucial in serving for the stability analysis \cite{Kase2023, Gannouji2022, Tsujikawa21}. Furthermore, under certain circumstances, this  kind of stability analysis provides us with a method to set constraints on a modified theory based on the inherence of its self-consistency \cite{Tsujikawa21}. 

In this paper, we investigate the stability of the 2nd-order Lagrangian of a subclass of Horndeski theory under the odd-parity gravitational perturbations. 
As the most general model of scalar-tensor theories, the Horndeski theories (and beyond) have drawn a lot attentions recently \cite{Kase2023,Kobayashi2021, Minamitsuji2023, Mironov2023, Yurika2023}. 
In Horndeski’s theory, the action contains a scalar field and a metric tensor field, which give rise to the metric and scalar field equations with no derivatives beyond the second order. In comparing to GR, Horndeski’s theory has the same symmetry including local Lorentz invariance and diffeomorphism \cite{houyu2023}. 
In fact, the stability problem in Horndeski theories has been studied intensively during the past decade \cite{Kobayashi2012, Minamitsuji2022}. Many typical configurations of the Lagrangian have been investigated. 
This time we shall focus on a set of specific background solutions of the theory and look into the details about how the stability is preserved or broken. As we have seen in \cite{Tsujikawa21}, the choice of coupling parameters at the Lagrangian level can affect the criterion for stability analysis in a comprehensive way. This is what we are going to demonstrate in here thanks to the analytic background solutions. 

The rest of the paper is organized as the following: Sec.\ref{secaction} provides the background hairy black hole solutions to the fields in Horndeski theory and demonstrates the odd-parity perturbations to that. Specially, a dramatically reduced 2nd-order Lagrangian is obtained from the original one. 
On top of that, we run the stability analysis in Sec.\ref{stab} and see how the instability emerges. Finally, some concluding remarks are given in Sec. \ref{conclusion}.

In this paper we are adopting the unit system so that $c=G=2M=1$, where $c$ denotes the speed of light, $G$ is the gravitational constant and $M$ stands for the total mass of the black hole.
All the Greek letter in indices run from 0 to 3. The other usages of indices will be described at suitable places.  


\section{Background fields and odd-parity gravitational perturbation}
\renewcommand{\theequation}{2.\arabic{equation}} \setcounter{equation}{0}
\label{secaction}

We consider an action in Horndeski theory with a scalar field denoted by $\phi(r)$ \cite{Bergliaffa2021}

\bqn
\lb{action1}
{\cal S} &=&  \int {d^4} x  \sqrt{-g} \big\{ Q_2(\chi) +Q_3(\chi) \Box \phi + Q_4(\chi) R \nb\\
&& +Q_{4, \chi} \big[  (\Box \phi)^2 -\left(\nabla^\mu \nabla^\nu \phi \right) \left(\nabla_\mu \nabla_\nu \phi \right) \big]  \big\} ,~~
\eqn
where
\bqn
\lb{action2}
{\chi} &\equiv& - \frac{1}{2} \nabla^\mu \phi \nabla_\mu \phi,
\eqn
and
\bqn
\lb{action3}
Q_2 &=& \alpha_{21} \chi + \alpha_{22} (-\chi)^{w_2}, \\
Q_3 &=& \alpha_{31} (-\chi)^{w_3}, \\
Q_4 &=& \kappa^{-2} + \alpha_{42} (-\chi)^{w_4},
\eqn
with $\kappa \equiv \sqrt{8 \pi}$. For the 4-current to vanish at infinity and the finiteness of energy, we require $\alpha_{21}=\alpha_{31}=0$ and $w_2=3 w_4=3/2$ \cite{Bergliaffa2021}.
The comma in the subscript stands for the derivative with respect to the quantity right after that, $R$ is the Ricci scalar, $\nabla$ denotes the covariant derivative operator, while the operator $\Box$ is defined as $\Box \equiv \nabla^{\mu} \nabla_{\mu} $. 

The static and spherically symmetric background metric is given in the line-element form as [in the Boyer-Lindquist coordinate $(t, r ,\theta, \phi)$]
\bqn
\lb{metric1}
ds^2 &=& -A dt^2 + B^{-1} dr^2 + r^2 d\Omega^2,
\eqn
where $d \Omega^2$ is the unit two-sphere line element. $A$ and $B$ are functions of $r$. Their explicit expressions, together with the background scalar field (denoted by $\phi_0$), are found to satisfy
\bqn
\lb{ABphi0}
A(r) = B(r) = 1-\frac{1}{r} +\frac{q}{r} \ln r, \nb\\
\phi_0' = \mathfrak{k} \frac{2}{r} \sqrt{-\frac{\alpha_{42}}{3 B \alpha_{22}}}, 
\eqn
where a prime in the superscript denotes the derivative with respective to $r$ and $\mathfrak{k}=\pm 1$.
The charge $q$ satisfies $q = (2/3)^{3/2} \kappa^2 \alpha_{42} \sqrt{-\alpha_{42}/\alpha_{22}}$. Clearly, we must require $\alpha_{42} \cdot \alpha_{22} \leq 0$. 

With the above background fields in hand, we are on the position to consider the odd-parity perturbations \footnote{\textcolor{black}{The ``odd-parity'', or equivalently ``axial'', gravitational perturbation got its name mainly due to the individual properties of parity transformation when decomposing as tensor harmonics \cite{Maggiore2018}. Basically, the odd-parity part will catch a factor of $(-1)^{l+1}$ under the parity transformation, which is different from its counterpart in the even-parity sector, viz., $(-1)^{l}$ \cite{Zerilli1970}. Here for simplicity, by following the notation system of, e.g., \cite{Thomp2017}, these odd-pairty parts are collected and given in the form of \eqref{hab}, as will be seen shortly.}} 
to them. Notice that, since scalar field perturbation only has even-parity contributions \cite{Kase2018, Ganguly2018}, here we only consider the gravitational perturbation for the odd-parity sector.  

We can spell out the metric as $g_{\mu \nu} = {\bar g}_{\mu \nu} + \epsilon h_{\mu \nu}$, where $\epsilon$ is a  bookkeeping parameter (in contrast, for the scalar field we simply have $\phi=\phi_0$).
The perturbation function $h_{\mu \nu}$ could be parameterized as \cite{Chao2023a}

\begin{widetext}
	\bqn
	\lb{hab}
	h_{\mu \nu} &=&  \sum_{l=0}^{\infty} \sum_{m=-l}^{l}
	\begin{pmatrix}
		0	& 0 &  C_{lm}\csc \theta \partial_\varphi & -C_{lm} \sin \theta \partial_\theta\\
		0	& 0 &  J_{lm}\csc \theta \partial_\varphi & -J_{lm} \sin \theta \partial_\theta\\
		sym	& sym &  G_{lm}\csc \theta \big(\cot \theta \partial_\varphi-\partial_\theta \partial_\varphi \big) &  sym \\
		sym	& sym &  \frac{1}{2} G_{lm}\big(\sin \theta \partial^2_\theta - \cos \theta \partial_\theta- \csc \theta \partial^2_\varphi\big) &   - G_{lm} \sin \theta \big(\cot \theta \partial_\phi-  \partial_\theta \partial_\varphi \big)
	\end{pmatrix}
	Y_{l m}(\theta, \varphi), 	\nb\\
\eqn
\end{widetext}
here  $C_{lm}$, $J_{lm}$ and $G_{lm}$ are functions of $t$ and $r$, while $Y_{l m}(\theta, \phi)$ denotes the spherical harmonics. Starting from now on, we shall set  $m=0$ in the above expressions so that $\partial_\phi Y_{lm}(\theta, \phi)=0$, as now the background has the spherical symmetry, and the corresponding linear perturbations do not depend on $m$   \cite{Regge57,Thomp2017}. 
In addition, we shall adopt the gauge condition ${G}_{lm}=0$ (This could be referred as the RW gauge \cite{Thomp2017}) in the following, which will bring $C_{lm}$ and $J_{lm}$ to gauge invariants \cite{Chao2023a}. 
For simplicity, in the following we shall drop the subscript ``$lm$'' before stimulating any confusions. 

By substituting the full metric and scalar field back into the Lagrangian [the integrant of the action \eqref{action1}], and picking up the ${\cal O} (\epsilon^2)$ terms, we obtain

\bqn
\lb{Lodd}
{\cal L}_{\text{odd}} &=& L\Big( \beta_1 {\dot J}^2 -2 \beta_1 {\dot J} {C^\prime} + \beta_1 \frac{4}{r}{\dot J} C\nb\\
&& + \beta_1 {C^{\prime 2}} + \beta_2 J^2 + \beta_3 C^2 \Big),  
\eqn
where
\begin{widetext}
\bqn
\lb{beta123}
\beta_1 &\equiv&  \frac{1 }{2 \kappa ^2} \sqrt{\frac{B}{A}}, \nb\\
\beta_2 &\equiv& \frac{A^2}{8 A^{3/2} \kappa ^2 r^2 \phi '}  \bigg[8 B^{3/2} \phi '+2 \sqrt{2} B^2 \kappa ^2 \left(4 \alpha _{42} \left(\phi '\right)^2+2 \alpha _{42} r^2 \left(\phi ''\right)^2+\left(\alpha _{22}-2 \alpha _{42}\right) r^2 \left(\phi '\right)^4-2 \alpha _{42} r^2 \phi ^{(3)} \phi '\right) \nb\\
&& ~~~~~ +\sqrt{2} \alpha _{42} \kappa ^2 r^2 \left(B'\right)^2 \left(\phi '\right)^2-2 \sqrt{2} \alpha _{42} B \kappa ^2 r \phi ' \left(r B'' \phi '+B' \left(r \phi ''+2 \phi '\right)\right)-4 \sqrt{B} L \phi '\bigg] \nb\\
&&+  \frac{2 \sqrt{2} \alpha _{42} B^2 \kappa ^2 r^2 \left(A'\right)^2 \left(\phi '\right)^2-A B r \phi ' \left(2 \sqrt{2} \alpha _{42} B \kappa ^2 r A'' \phi '+\sqrt{2} \alpha _{42} \kappa ^2 r A' B' \phi '-8 \sqrt{B} A'\right)}{8 A^{3/2} \kappa ^2 r^2 \phi '}, \nb\\
\beta_3 &\equiv& \frac{A^2 }{16 A^{5/2} B \kappa ^2 r^2 \phi '} \bigg\{ 2 \sqrt{2} B^2 \kappa ^2 r \left(2 \alpha _{42} r \left(\phi ''\right)^2+\left(\alpha _{22}-2 \alpha _{42}\right) r \left(\phi '\right)^4-4 \alpha _{42} \phi ' \left(r \phi ^{(3)}+2 \phi ''\right)\right) \nb\\
&& ~~~~~ +8 \sqrt{B} \phi ' \left(L-r B'\right)+\sqrt{2} \alpha _{42} \kappa ^2 r^2 \left(B'\right)^2 \left(\phi '\right)^2 \nb\\
&& ~~~~~ +4 \sqrt{2} \alpha _{42} B \kappa ^2 \phi ' \left[\phi ' \left(r^2 \left(-B''\right)-4 r B'+L\right)-2 r^2 B' \phi ''\right]\bigg\} \nb\\
&& +\frac{3 \sqrt{2} \alpha _{42} B^2 \kappa ^2 r^2 A^{\prime 2} \phi^{\prime 2}-4 A B r \phi ' \Big[ \sqrt{2} \alpha _{42} \kappa ^2 r A' B' \phi '+2 \sqrt{B} A'+\sqrt{2} \alpha _{42} B \kappa ^2 \left(r A'' \phi '+A' \left(r \phi ''+2 \phi '\right)\right) \Big]}{16 A^{5/2} B \kappa ^2 r^2 \phi '}, \nb\\
\eqn
\end{widetext}
 $L\equiv l (l+1)$, and a dot over the head stands for the time derivative. The ``$(n)$'' in the superscript means the nth derivative with respect to $r$.
Notice that, in obtaining \eqref{Lodd}, integration by parts \cite{Arfken} and the properties of spherical harmonics \cite{Chao2023a} have been used continuously. 
Also notice that, the above Lagrangian could reduce to that of GR at the $\alpha_{42}=\alpha_{22}=0$ limit. 

By mimicking \cite{Tsujikawa21} and introducing a new gauge invariant
\bqn
\lb{zeta}
\zeta &\equiv& \frac{2}{r} C-C^\prime+{\dot J},
\eqn
the Lagrangian \eqref{Lodd} becomes
\bqn
\lb{Lodd2}
{\cal L}_{\text{odd}} &=& L\bigg\{ \beta_1 \left[\zeta^2 -2 \zeta \left( \frac{2}{r} C-C^\prime+{\dot J} \right) \right]\nb\\
&&  + \beta_2 J^2 + \left( \beta_3-\frac{2}{r^2} \beta_1-\frac{2}{r} \beta_1^\prime \right) C^2 \bigg\},~~~ 
\eqn
for which the Euler-Lagrange (E-L) equation \cite{Taylor05} could be applied on $C$ and $J$ so that their expressions in terms of $\zeta$ could be solved for. As we have done in, e.g.,  \cite{Chao2023a} and \cite{Tsujikawa21}, these solved expressions could be substituted back into the \eqref{Lodd2} and lead to a Lagrangian solely of one variable $\zeta$ (since non-dynamical terms have been eliminated). Such a Lagrangian could be written as
\bqn
\lb{Lodd3}
{\cal L}_{\text{odd}} &=& \mathbb{K} {\dot \zeta}^2+\mathbb{G} { \zeta}^{\prime 2}+\mathbb{N} { \zeta}^2.
\eqn
Due to their tediousness, we abbreviate the full expressions of $\mathbb{K} $, $\mathbb{G} $ and $\mathbb{N} $ in here. For those who need them, please check the supplemental material in \cite{supplemental1}. Once again, the  Lagrangian \eqref{Lodd3} reduces to that of GR at the $\alpha_{42}=\alpha_{22}=0$ limit. 


\section{Stability analysis}
\renewcommand{\theequation}{3.\arabic{equation}} \setcounter{equation}{0}
\label{stab}

Let us work on top of the reduced Lagrangian \eqref{Lodd3}. 
According to \cite{Kase2018}, the no-ghost stability condition requires $\mathbb{K}>0$. Using \eqref{ABphi0} on $\mathbb{K}$, at the point $r=1$ it becomes
\bqn
\lb{K11}
\left. \mathbb{K} \right|_{r\to1}&=& \frac{81 \sqrt{\frac{3}{2}} \alpha _{22} \sqrt{-\frac{\alpha _{42}}{\alpha _{22}}} L}{64 \pi ^2 \alpha _{42}^2 \left(16 \pi  \sqrt{6} \sqrt{-\frac{\alpha _{42}}{\alpha _{22}}} \alpha _{42}+9\right){}^2 \mathfrak{k}}.\nb\\
\eqn
Clearly, the no-ghost condition holds only when ${\alpha _{22}} \cdot \mathfrak{k} > 0$, which implies $-{\alpha _{42}} \cdot \mathfrak{k} > 0$. 
For later convenience, let us introduce a set of reduced coupling parameters (which are always positive) by $\kappa_{22}\equiv \alpha_{22}/\mathfrak{k}$ and  $\kappa_{42}\equiv -\alpha_{42}/\mathfrak{k}$. 

By referring \cite{Tsujikawa21}, we can define the propagation speed at the radial direction by $c_r \equiv d r_\ast/dt={\hat c}_r/A$. Here ${\hat c}_r$ is a quantity introduced to better describe the radial Laplacian stability condition, which is found to satisfy ${\hat c}_r^2 \mathbb{K} + \mathbb{G} = 0$ \cite{Tsujikawa21}, so that \textcolor{black}{ ${\hat c}_r^2=-\mathbb{G} \cdot \mathbb{K}^{-1}$.} On top of that, the radial Laplacian stability condition is  given by ${\hat c}_r^2 \geq 0$.

\begin{figure}[htb]
	\includegraphics[width=\columnwidth]{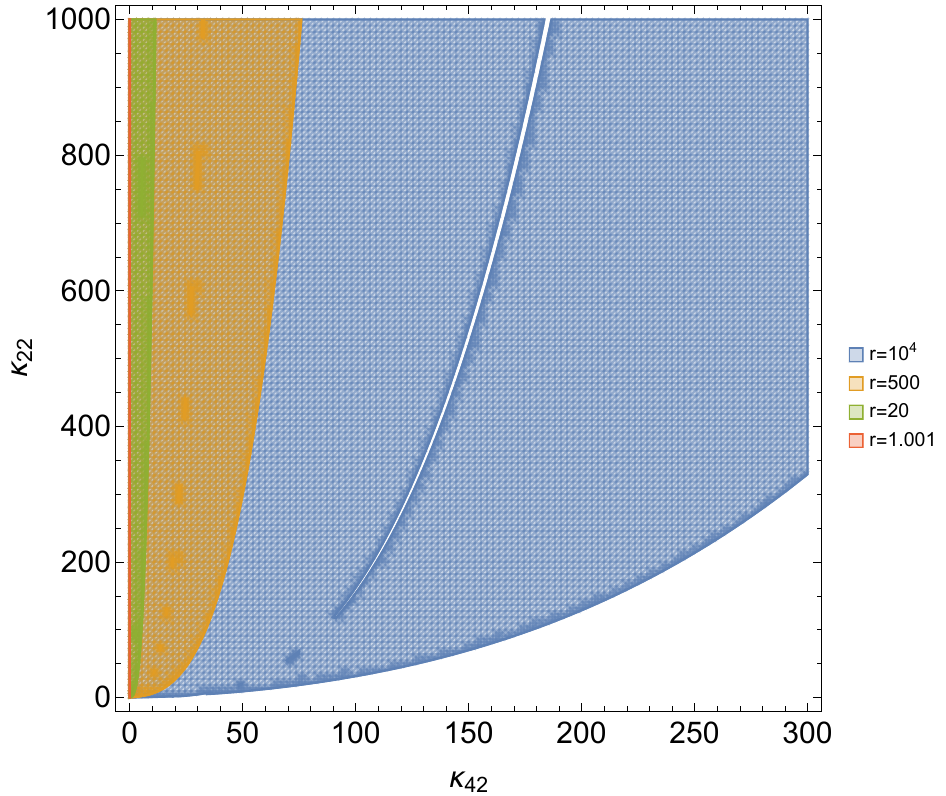} 
 	\includegraphics[width=\columnwidth]{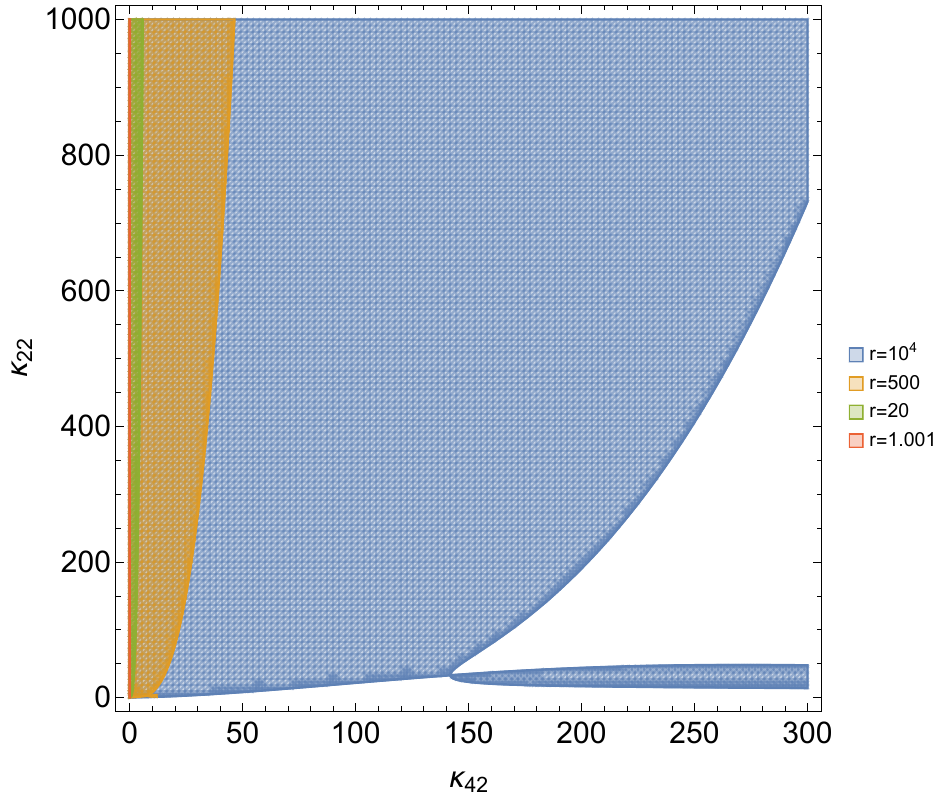} 
\caption{The phase space of $(\kappa_{42}, \kappa_{22})$ for which the shadowed regions mean ${\hat c}_r^2 \geq 0$, i.e., the radial Laplacian condition gets satisfied. The blue, orange, green and red shadowed regions correspond to $r=10^4,\;500,\;20,\;1.001$ respectively. Upper panel: $\mathfrak{k} = 1$; lower panel: $\mathfrak{k} = -1$. Here we have set $l=2$. Notice that, in here only the first quadrant of $(\kappa_{42}, \kappa_{22})$ are considered since $\kappa_{42}$ and $\kappa_{22}$ are defined to be positive (cf., Sec.\ref{stab}).} 
	\label{plot1}
\end{figure} 

To monitor the behavior of ${\hat c}_r^2$ as a function of $r$, we plot it out in the phase space of $(\kappa_{42}, \kappa_{22})$ by setting $l=2$ in Fig. \ref{plot1} for various $r$'s, in which both $\mathfrak{k} = \pm 1$ cases are considered.  At those colorful shadowed regions the radial Laplacian condition, viz., ${\hat c}_r^2 \geq 0$, gets satisfied. In there the quantity ${\hat c}_r^2$ is considered at different positions by varying $r$. We observe from Fig. \ref{plot1} that the "stable region" is shrinking as $r$ approaching the event horizon $r=1$.  
What is omitted in Fig. \ref{plot1} is that the stable region will disappear if $r$ is sufficiently close to $1$, e.g., when $r=1+ 10^{-20}$. That means we definitely have this kind of instability no matter how the cupling parameters are chosen.

To make it more manifest, let us consider the quantity ${\hat c}_r^2$ in the neighborhood of event horizon $r=1$ \textcolor{black}{(without setting $l$ to a specific value)} and insert the full expressions \cite{supplemental1}. That leads to
\bqn
\lb{G11}
\left. {\hat c}_r^2 \right|_{r \to 1} &=& -\frac{2}{81}  (r-1)^2 \left(16 \sqrt{6} \pi  \kappa _{42} \sqrt{\frac{\kappa _{42}}{\kappa _{22}}} \mathfrak{k}-9\right)^2 \nb\\
&& + {\cal O}(r-1)^3.
\eqn
Thus, for any combination of the coupling parameters, the radial Laplacian instability always exists in the neighborhood of $r=1$, as we mentioned above.  
Such a result is actually consistent with the conclusions given in \cite{Minamitsuji2022, Kobayashi2012, Kobayashi2014} \footnote{The conclusions of \cite{Kobayashi2012, Kobayashi2014} are questioned by \cite{Babichev2018}, although (to the best of our knowledge) the relative arguments have not been fully taken into consideration yet by the literature \cite{Kobayashi2019}. With a more comprehensive analysis to Horndeski models by referring \cite{Babichev2018}, it is possible for one to find clues for their survival. However, so far the analysis to the model considered in here seems to be concrete. We shall expediently ignore the controversy arisen by \cite{Babichev2018}. }. 


\section{Conclusions}
\renewcommand{\theequation}{4.\arabic{equation}} \setcounter{equation}{0}
\label{conclusion}

In this paper we focus on a specific subclass of Horndeski theory describing by the action \eqref{action1}. 
Thank to a set of analytic background solutions \eqref{ABphi0} (with their deviations from that of GR are mainly characterized by two coupling parameters), we are able to systematically investigate the odd-parity gravitational perturbation to that background and extract the 2nd-order Lagrangian of the theory.  With suitable mathematical techniques, the original Lagrangian finally reduced to the form of \eqref{Lodd3}, with just one degree of freedom as expected\footnote{Notice that, due to the tediousness of the factors of  \eqref{Lodd3}, their explicit expressions are omitted in here and provided in \cite{supplemental1} instead. }. On top of that, we further run the stability analysis (see., e.g., \cite{Tsujikawa21, Kase2023, Kase2018}) and demonstrate how the instability emerges.  

First of all, certain constraints to the coupling parameters are identified from the no-ghost condition [cf., \eqref{K11}]. Based on that, the following calculations can be simplified. This enables us to calculate and analyze the radial Laplacian stability condition within the valid phase space of (reduced) coupling parameters (cf., Fig.\ref{plot1}). In Fig.\ref{plot1} we learn that such a stability condition requires additional constraints to the coupling parameters. More importantly, this stability condition tends to be broken in the neighborhood of event horizon $r=1$. Indeed, by using the asymptotic expansion of the criterion ${\hat c}_r^2$ [cf., \eqref{G11}], it is clear that this condition will not be preserved anymore as $r \to 1$ [although we also notice that such a condition can be satisfied right at the point $r=1$, since we precisely have ${\hat c}_r^2(r=1)=0$] no matter how the (physical) coupling parameters are chosen. We thus conclude that the set of background solutions  \eqref{ABphi0} are not stable and needs to be excluded from the valid solutions, which is consistent with the predictions  given in \cite{Minamitsuji2022}.

As observed from different occasions, setting those coupling parameters to zero (as in the GR limit) in the final  criterion or discriminant [e.g., the \eqref{G11} in this paper]  will not always simply bring us back to that of GR, different from what we have seen at the Lagrangian level. 
In \cite{Tsujikawa21, Kobayashi2014} we also observed similar phenomena.  This implies that (in)stability of a theory could be a comprehensive effect and it is not necessary for the  criterion to behave like a perturbation to the stable GR case. 
It is also interesting to mention that, as we see from the Fig.\ref{plot1}, the most stringent constraint to the coupling parameters are given by the analysis within the neighborhood of $r=1$ (instead of the other positions that far from the event horizon). That is similar to the situation shown in \cite{Kase2018}. 

In the future, we can move to other background solutions and models of the Horndeski theory. By referring \cite{Minamitsuji2022}, we can pay more attention to those stable cases. In principle, the corresponding stability analysis can not only help us bring constraints to the model but also give a reduced Lagrangian as a byproduct, which can serve for the QNM calculations in the next step. 
With more models investigated, it is also possible to extract some common features by using the conclusions from different cases.


\section*{Acknowledgments}

We would like to express our gratitude to Prof.  Anzhong Wang for his valuable comments and suggestions. This work is supported in part by the National Key Research and Development Program of China Grant No. 2020YFC2201503, the National Natural Science Foundation of China under Grant No. 12205254, 12275238,  and 11675143,  the Zhejiang Provincial Natural Science Foundation of China under Grant No. LR21A050001 and LY20A050002, the Fundamental Research Funds for the Provincial Universities of Zhejiang in China under Grant No. RF-A2019015, and the Science Foundation of China University of Petroleum, Beijing under Grant No. 2462024BJRC005.




\end{document}